\documentclass[aps,prb,twocolumn,superscriptaddress,groupedaddress]{revtex4}  % for review and submission
\usepackage{graphicx}
\usepackage{amsmath}
\usepackage{float}
\usepackage{color}

\newcommand {\NF}{{N_{\rm F}}}

\newcommand {\bbk}{{\bf k}}
\newcommand {\bbq}{{\bf q}}
\newcommand {\bbkk}{{\bf {k'}}}
\newcommand {\ek}{{\epsilon_{\bf k}}}

\newcommand {\oql}{{\omega_{\bf q \nu}}}
\newcommand {\on}{{\omega_n}}
\newcommand {\onp}{{\omega_{n'}}}
\newcommand {\afo}{{\alpha^{2} F(\omega)}}
\newcommand {\gkkl}{{g_{\bf k \bf {k'}}^{\nu}}}
\newcommand {\deltak}{{\delta(\epsilon_{\bf k})}}
\newcommand {\deltakk}{{\delta(\epsilon_{\bf {k'}})}}

\begin{document}

\title{Two-gap superconductivity in heavily \textit{n}-doped graphene: \\
\textit{ab initio} Migdal-Eliashberg theory}

\author{E. R. Margine}
\email{rmargine@binghamton.edu}
\affiliation{Department of Physics, Applied Physics and Astronomy, 
Binghamton University - SUNY, Binghamton, New York 13902, USA}
  
\author{Feliciano Giustino}
\email{feliciano.giustino@materials.ox.ac.uk}
\affiliation{Department of Materials, University of Oxford, Parks Road,
Oxford OX1 3PH, United Kingdom}

\begin{abstract}
Graphene is the only member of the carbon family from zero- to three-dimensional
materials for which superconductivity has not been observed yet.
At this time, it is not clear whether the quest for superconducting graphene is hindered by technical challenges, or else by the fluctuation of the order parameter in two dimensions.
In this area, {\it ab initio} calculations are useful to guide experimental efforts by narrowing down the search space. In this spirit, we investigate from first principles the possibility of inducing superconductivity in doped graphene using the fully anisotropic Migdal-Eliashberg theory powered by Wannier-Fourier interpolation.
To address a best-case scenario, we consider both electron 
and hole doping at high carrier densities, so as to align the
Fermi level to a van Hove singularity.
% 0.6 el / (sqrt(3)/2*a^2) with a = 2.434 angstrom gives 11.69 * 10^14 cm-2 
In these conditions, we find superconducting gaps of $s-$wave symmetry,
with a slight anisotropy induced by the trigonal warping, and, in the case of 
$n$-doped graphene, an unexpected two-gap structure 
reminiscent of MgB$_2$. Our Migdal-Eliashberg 
calculations suggest that the observation
of superconductivity at low temperature should be possible for $n$-doped graphene at carrier 
densities exceeding $10^{15} \mbox{cm}^{-2}$.
\end{abstract}

\maketitle

\section{INTRODUCTION}
%\noindent
Superconductivity in lightweight carbon-based materials was first discovered almost half a century ago
in alkali-metal doped graphite \cite{hannay}. Since then, the intercalation of metal atoms into graphite 
and fullerene solids has led to superconducting critical temperatures $T_c$ above 11~K in CaC$_6$ 
at ambient pressure \cite{weller} and 38~K in Cs$_3$C$_{60}$ under applied pressure \cite{ganin}. 
Substitutional doping of diamond with boron also induces a superconducting state, with a critical
temperature in the range 4-10~K~\cite{ekimov,Takano2007}.
The most recent breakthroughs within the family of carbon-based materials are the discoveries
of superconductivity in doped polycyclic aromatic hydrocarbons, namely picene~\cite{mitsuhashi}, 
phenanthrene~\cite{wang}, coronene~\cite{kubozono}, and dibenzopentacene~\cite{xue},
with critical temperatures up to 33~K~\cite{xue}.

The discovery of graphene~\cite{novoselov,novoselov-pnas} together with 
its unique properties~\cite{neto,bonaccorso} immediately raised the question of whether superconductivity could be achieved also in this two-dimensional material. Within this context,
theoretical studies explored both conventional and unconventional pairing mechanisms~\cite{einenkel,profeta,si,uchoa,nandkishore,kiesel}. In the former case, it was suggested that phonon-mediated superconductivity 
could be induced by tailoring the electron-phonon coupling (EPC) via alkali-metal doping~\cite{profeta}. 
In the latter case, it was proposed that chiral 
superconductivity should arise when graphene is doped near the van Hove singularity (VHS), as a result of
strong electron-electron interactions \cite{nandkishore,kiesel}. 
Chemical modifications of graphene were also predicted to lead to superconductivity. For example, 
in close analogy to B-doped diamond \cite{ekimov}, hole-doped graphane \cite{sofo} was predicted 
to be a high-$T_c$ superconductor~\cite{savini}. 

Despite such a variety of theoretical predictions, so far none has been confirmed experimentally.
This raises the question of whether there exists a fundamental limitation preventing a 
superconducting phase transition in graphene, similar to the Mermin-Wagner theorem \cite{mermin}, or if
inducing and observing superconductivity is indeed possible but technically very challenging.

On the experimental front, the study of possible pairing mechanisms is complicated by the sensitivity of the EPC to the  character and location of the metal atoms, as well as the underlying substrates~\cite{mcchesney,bianchi,siegel,haberer}. For example, the EPC strength determined by angle-resolved photoelectron spectroscopy differs substantially in the cases 
of subsurface intercalation of potassium on graphene/Au~\cite{haberer} and potassium adsorption on graphene/Ir~\cite{bianchi}. 
This sensitivity may reflect the substrate-induced modification of the electronic structure in proximity of the Dirac point:
for example, graphene on Ir or Cu exhibits a
band gap~\cite{bianchi,siegel}, graphene grown on Au is gapless~\cite{haberer}, and in the case of SiC both situations
have been reported~\cite{mcchesney,zhou}.

Given the lack of experimental confirmation of current theories, and the difficulty in extracting
the relevant pairing parameters from experiment, in order to understand the potential of graphene 
for superconductivity it is important to carry out careful investigations using the most advanced 
tools available.

\begin{figure*}
\begin{center}
\includegraphics[width=\textwidth]{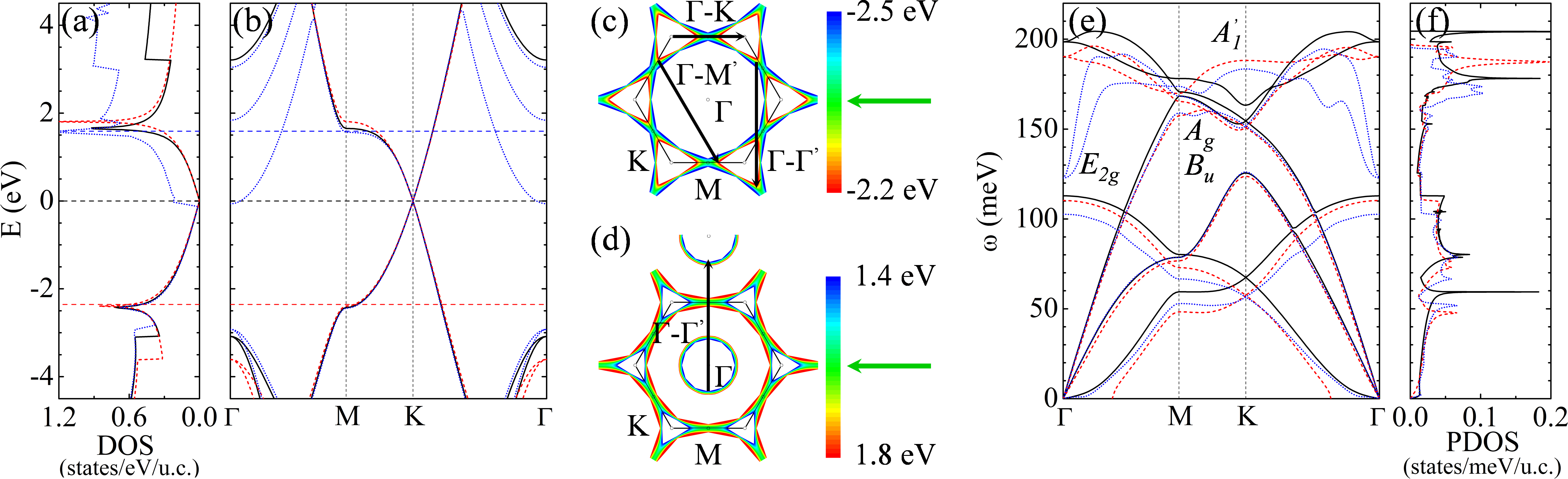}
\end{center}
\caption{ (Color online)
Electronic-structure and phonon-dispersion relations of pristine and doped graphene.
(a) Electronic density of states and (b) band structures of pristine graphene (black solid line), $p$-doped graphene (dashed red line), and $n$-doped graphene (dotted blue line).
(c),(d) Fermi surfaces of $p$-doped and $n$-doped graphene, respectively.
In both cases, the Fermi surface corresponds to the green region of the energy isosurface, as
indicated by the small arrow in the colorbar. The long black
arrows indicate the dominant electron-phonon scattering mechanisms at the Fermi surface.
(e) Phonon-dispersion relations of pristine, $p$-doped, and $n$-doped graphene,
using the same color code as in (a). The vibrational modes discussed in the text are indicated. (f) Phonon density of states corresponding to the dispersions in (e).
}
\label{fig1}
\end{figure*}

In this work we study from first principles the possibility of superconductivity in heavily doped 
graphene. We employ a recent {\it ab initio} implementation of the anisotropic 
Migdal-Eliashberg theory~\cite{allen_mitrovic,margine_eliashberg} based on electron-phonon
Wannier-Fourier interpolation~\cite{giustino_wannier,marzari_rmp}. This technique allows us to describe
the electron-phonon pairing mechanism by taking fully into account the highly anisotropic 
nature of graphene, and by sampling the Brillouin zone with unprecedented accuracy.
By considering high carrier densities, whereby the Fermi energy approaches 
a VHS, we show below that superconductivity should be atteinable at least in
$n$-doped graphene, due to the additional pairing channels associated with the 
free-electron-like (FEL) band.

\section{MODELING DOPED GRAPHENE AT HIGH CARRIER DENSITY}
%\noindent
In conventional superconductors the critical temperature increases with the electronic density of states at the Fermi level. In the case of graphene, this means that a best-case scenario for superconductivity should correspond to situations in which the Fermi level matches a VHS, either in the valence or in the conduction band. From our calculations, we find that the carrier
densities required to approach the VHS  are $0.8\cdot 10^{15}$~cm$^{-2}$ for $p$-doped graphene, and $1.2\cdot 10^{15}$~cm$^{-2}$ for $n$-doped graphene, respectively.
%in graphene are of the order of $10^{15}$~cm$^{-2}$. 
Carrier densities  near the VHS have been achieved via two-sided alkali metal doping of graphene~\cite{mcchesney}, 
% max doping in their case corresponds to 2 K atoms per hexagon (up and down), ie 4E5 cm-2
 and densities up to $4\cdot 10^{14}$~cm$^{-2}$ via electrolytic gating~\cite{efetov}. While carrier densities up to $3.5\cdot 10^{15}$~cm$^{-2}$ have been demonstrated recently 
by means of polymer/electrolyte gating of gold thin films~\cite{gonnelli}, such values may not be easily attainable in graphene due to its quantum capacitance~\cite{Ye_PNAS11,Uesugi_SREP12}.

Here we simulate carrier doping using a jellium model, whereby the excess/defect electronic charge is compensated by a uniform neutralizing background. This approximation has been used in previous studies to examine the effect of doping on the electronic bands, phonon dispersions, and EPC strength in carbon nanotubes~\cite{margine_nfe,margine_phonons}, graphite intercalation compounds~\cite{csanyi,boeri_gic}, and graphene~\cite{si,piscanec,lazzeri_nonadiab,Park1,Park2,Calandra2007,giustino_eph}. Besides being computationally advantageous, the jellium model is expected to provide a realistic description of graphene doped via electrochemical gating~\cite{efetov,gonnelli}. 
In fact, while doping is expected to modify the band structure in proximity of the Dirac point~\cite{craciun,lui}, at high carrier densities its effects on the Fermi surface and the EPC are expected not to be significant. 
In the related case of superconducting gated MoS$_2$~\cite{Ye_MoS2_supercond}, first-principles calculations using the same jellium model were able to reproduce the measured trends in the superconducting transition temperature~\cite{Ge_MoS2_supercond}.
The calculations were carried out within the local density approximation to density-functional theory, using the codes {\tt Quantum-ESPRESSO}~\cite{QE}, {\tt EPW}~\cite{EPW}, and {\tt Wannier90}~\cite{wannier}. The technical details of these calculations are described in Sec.~\ref{methodology} at the end of this paper.

\begin{figure*}
\centering
\includegraphics[width=\textwidth]{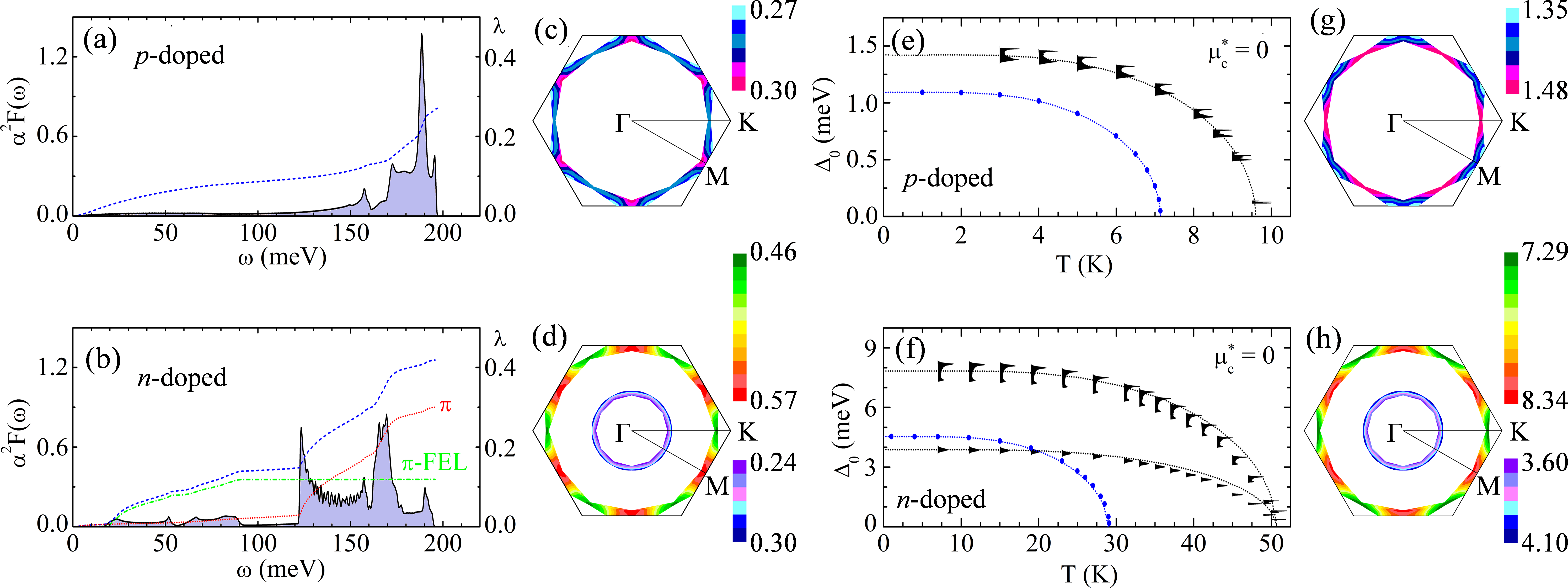}
\caption{ (Color online)
Electron-phonon coupling and superconducting gap function in doped graphene.
(a), (b) Eliashberg spectral function and cumulative EPC calculated for $p$-doped and $n$-doped graphene,
respectively. $\alpha^2F$ is the black solid line (left scale), $\lambda(\omega)$ is the blue dashed line
(right scale). In the case of $n$-doped graphene we also show the decomposition of $\lambda(\omega)$ into
contributions from the intraband $\pi^*$ scattering (red dotted line) and interband $\pi^*$-FEL scattering (green dash-dotted line).
(c), (d) Momentum-resolved EPC $\lambda_{\bbk}$ for electrons on the Fermi surface, for $p$-doped and $n$-doped graphene, respectively.
(e)-(h) Anisotropic superconducting gap of $p$-doped graphene
and $n$-doped graphene calculated using the Migdal-Eliashberg equations and $\mu^*_c=0$.
Panels (e) and (f) represent the energy distribution of the gap across
the Fermi surface at various temperatures (shown in black), for $p$-doped and $n$-doped graphene,
respectively. The isotropic superconducting gap is shown as blue filled dots.
The dotted lines are BCS fits to the calculated data.
Panels (g) and (h) show the corresponding superconducting gap at zero temperature on the Fermi surface (in meV). In the case of $n$-doped graphene, there are two superconducting gaps, one for the $\pi^*$ band and one for the FEL band (h).
}
\label{fig2}
\end{figure*}

Figure~\ref{fig1} shows the calculated band structures, Fermi surfaces, and phonon-dispersion relations of pristine as well as doped graphene. 
In analogy with the interlayer state of graphite intercalation compounds, the FEL band crosses the Fermi level for $n$-doped graphene as shown in Fig.~\ref{fig1}(b) (blue dotted lines). 
Upon doping, the highest optical phonon branches corresponding to the 
$E_{2g}$ modes at ${\it \Gamma}$ and the $B_{3u}$, $B_{2u}$, and $A_g$ modes at $M$ become softer, 
while the $A_{1}'$ mode at $K$ hardens [Fig.~\ref{fig1}(e)]. 
This behavior can be rationalized in terms of the changes 
intervening in the Fermi surface upon doping.
In fact, it is well known that in pristine 
graphene the scattering of electrons around ${\it \Gamma}$ 
and between ${\it \Gamma}$ and $K$ leads to two Kohn anomalies 
in the phonon dispersions at ${\it \Gamma}$ and $K$~\cite{piscanec}. When graphene is doped until the Fermi level matches one of the VHS [Fig.~\ref{fig1}(b)], 
the topology of the Fermi surface is dramatically
altered and we obtain a hexagon connecting adjacent $M$ points [Figs.~\ref{fig1}(c) and ~\ref{fig1}(d)]. 
As a result, the ${\it \Gamma}$ to $K$ scattering channel is suppressed, while scattering between flat parallel sheets of the Fermi surface becomes possible, 
as shown schematically in Fig.~\ref{fig1}(c) and (d). 
As a consequence, the original Kohn anomaly at $K$ is lifted, and significant
phonon softening is observed at ${\it \Gamma}$ and $M$, corresponding with ${\it \Gamma}$-${\it \Gamma'}$ and ${\it \Gamma}$-$M'$ scattering across adjacent Brillouin zones, respectively. 
We also note that the softening of the lowest acoustic branch 
in $p$-doped graphene leads to a dynamical instability. This instability is expected to be mitigated by the coupling with a substrate; in any event, the soft mode does not contribute to the electron-phonon coupling in the following analysis.

In principle, the electron-hole symmetry around the Dirac point should
result into similar Fermi surfaces for $p$-doping and $n$-doping. However, in the latter case the presence of a FEL band introduces an additional Fermi-surface sheet centered at ${\it \Gamma}$, as shown in Fig.~\ref{fig1}(d).
This extra sheet opens an additional ${\it \Gamma}$-${\it \Gamma'}$ scattering channel, and it is responsible
for the more pronounced phonon softening near ${\it \Gamma}$ in the case of $n$-doping.
The details of the phonon softening/hardening upon doping are expected 
to change slightly when the lattice constant of doped graphene, non-adiabatic 
corrections~\cite{lazzeri_nonadiab}, or the presence of dopant atoms are explicitly taken into account~\cite{profeta}. Nevertheless the general trends should be insensitive to these effects.

\section{ELECTRON-PHONON INTERACTION IN HEAVILY DOPED GRAPHENE}
%\noindent
Within the framework of conventional superconductivity, Cooper pairing arises from the interaction between electrons and phonons. In order to analyze the strength of this interaction, in Fig.~\ref{fig2} we show the isotropic Eliashberg spectral function, $\afo$, the cumulative EPC, $\lambda(\omega)$, and the momenum-resolved EPC of each electronic state at the Fermi surface, $\lambda_{\bbk}$. Here $\omega$ and $\bbk$ represent
the vibrational frequency and the electron momentum, respectively, and explicit expressions for $\afo$, $\lambda(\omega)$, and $\lambda_{\bbk}$ are provided in Sec.~\ref{methodology}.

\begin{figure*}[t!]
\centering
\includegraphics[width=\textwidth]{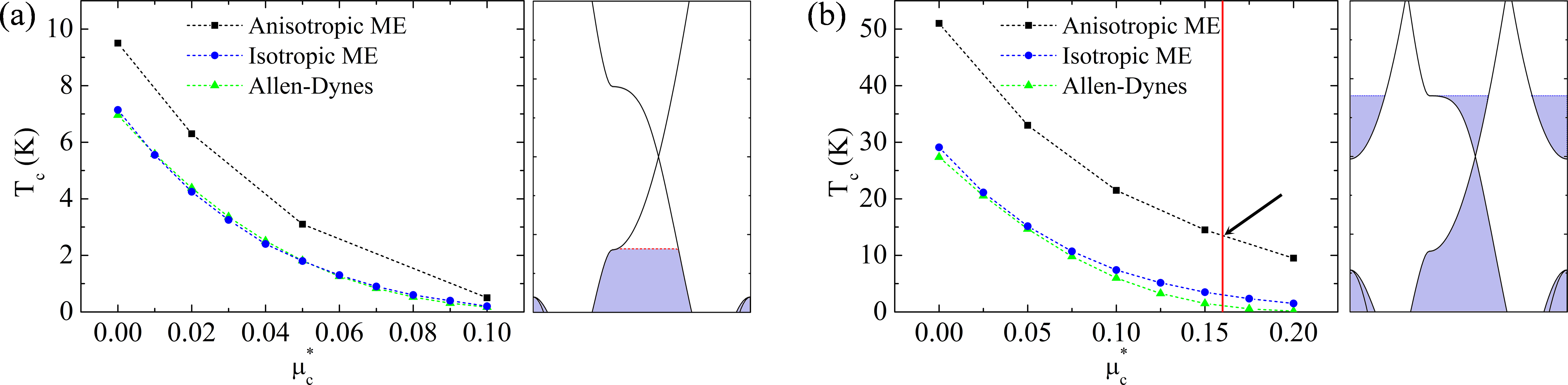}
\caption{ (Color online)
Superconducting critical temperature in doped graphene vs.\ Coulomb parameter.
(a), (b) Calculated superconducting critical temperature as a function
of the Coulomb pseudopotential $\mu^*_c$, for $p$-doped graphene and $n$-doped graphene, respectively.
The filled black squares are from anisotropic Migdal-Eliashberg calculations,
the filled blue circles correspond to the isotropic approximation, and
the filled green triangles are estimates based on the Allen-Dynes equation.
The lines are guides to the eye.
The vertical red line in (b) indicates the Coulomb parameter estimated
here using the model dielectric function of Ref.~\protect\cite{DasSarma2007},
%Ref.~\protect\onlinecite{DasSarma2007},
and the small arrow indicates our best estimate for the $T_c$ of $n$-doped graphene. The schematic band-structure plots illustrate the two doping scenarios considered in this work.
}
\label{fig3}
\end{figure*}

Starting from $p$-doped graphene, $\lambda(\omega)$ displays a smooth increase as a function of phonon frequency, and the sharp peaks in $\afo$ above 140~meV, corresponding to the optical modes, account for more than
half of the total EPC, $\lambda=0.27$ [Fig.~\ref{fig2}(a)].
Fig.~\ref{fig2}(c) shows that the momentum-resolved EPC is rather uniform 
across the Fermi surface, with $\lambda_{\bbk}$ varying smoothly in the range 0.27-0.30.

In the case of $n$-doped graphene, shown in  Fig.~\ref{fig2}(b), the Eliashberg function exhibits additional structure in the energy range 120-180~meV. By direct comparison with Fig.~\ref{fig1}(e)
this structure can be assigned to the large Kohn anomaly at ${\it \Gamma}$ induced by doping.
The softer phonons and the higher density of states at the Fermi level
arising from the additional FEL band (1.01 for $n$-type doping
vs.\ 0.62~states/eV/cell in the case of $p$-type doping)
lead to a substantial increase of the total EPC with respect to $p$-doped graphene, up to $\lambda=0.42$. In this case the optical phonons above 90~meV contribute almost two-thirds 
of the total~EPC.

To clarify the role of the FEL band in $n$-doped graphene we show in Fig.~\ref{fig2}(b) the contributions to $\lambda(\omega)$ associated with intraband scattering within the $\pi^*$ sheet, and interband scattering between the $\pi^*$ and FEL sheets.
From this plot and the decomposition of $\lambda(\omega)$ into the contributions from each phonon mode, we deduce that the optical in-plane C-C stretching phonons are responsible for intraband scattering in the $\pi^*$ sheet, while the out-of-plane buckling modes give rise to $\pi^*$-FEL interband scattering. The intraband scattering within the FEL sheet is found to be negligible. This phenomenology is similar to that of graphite-intercalation compounds~\cite{boeri_gic}, however in the present case the $\pi^*$-FEL coupling is much smaller, due to the fact that the FEL wavefunction is located farther away from the graphene layer~\cite{profeta}. In agreement with these observations, Fig.~\ref{fig2}(d) shows that the momentum-resolved coupling $\lambda_{\bbk}$ is relatively uniform on each Fermi surface sheet, but it differs considerably between the $\pi^*$ sheet (0.46-0.57) and the FEL sheet (0.24-0.30).

\section{SUPERCONDUCTING PAIRING}
%\noindent
Having examined the EPC in doped graphene we now move on to investigate its effect on the superconducting pairing. We solve the anisotropic Migdal-Eliashberg (ME) equations~\cite{allen_mitrovic, choi1, choi2} using the method described in Ref.~\cite{margine_eliashberg}.
%Ref.~\onlinecite{margine_eliashberg}
The solution to the ME equations, which are given explicitly in Sec.~\ref{methodology}, provides the complete superconducting gap function $\Delta(\bbk,\omega)$ across the Fermi surface. From this quantity, the leading edge of the superconducting gap for electron momenta $\bbk$ on the Fermi surface is obtained by solving for $\omega$ in $\Re\,\Delta(\bbk,\omega)=\omega$~\cite{margine_eliashberg}.

Since in the ME theory the Coulomb repulsion between electrons typically counters the electron-phonon pairing, as a best-case scenario we consider first the case of vanishing Coulomb pseudopotential, $\mu_c^*=0$ (see Sec.~\ref{methodology}). Figure~\ref{fig2} shows the calculated distribution of the leading edge $\Delta_0$ of the superconducting gap.
In the case of $p$-doped graphene, the gap function is slightly anisotropic
[anisotropy ratio $(\Delta_0^{\rm max}-\Delta_0^{\rm min})/\Delta_0^{\rm ave}=11\%$], however the symmetry is clearly $s$-wave as can be seen in Fig.~\ref{fig2}(g). For a carrier density corresponding to 0.4~holes/cell we obtain a critical temperature $T_c(\mu_c^*=0)=9.5$~K and a ratio $2 \Delta_0 / k_B T_c = 3.44$, very close to the ideal BCS value of 3.53. 
The temperature dependence of the superconducting gap is well described by
a BCS model, as obtained by solving numerically the BCS gap equation~\cite{BCS} using $\Delta_0$ and $T_c$ from our first-principles calculations. This is shown by the dotted black line in Fig.~\ref{fig2}(e).
%\begin{eqnarray} 
%\frac{1}{\lambda} \approx \int_0^{\delta^{-1} \sinh(\lambda^{-1})}
%                { \frac{dz}{1+z^2} \tanh(0.882 \frac{\delta}{\tau} \sqrt{1+z^2})},
%\label{BCS_gap}   
%\end{eqnarray}
%where $\delta=\Delta_0(T)/\Delta_0(T=0K)$ and $\tau=T/T_c$.

In the case of $n$-doped graphene, the presence of two Fermi surface sheets leads naturally to a two-gap structure, similar to the case of MgB$_2$~\cite{Liu2001,choi1,choi2}. As shown by Figs.~\ref{fig2}(f) and ~\ref{fig2}(h), also in this case the gaps have $s$-wave symmetry and are slightly anisotropic. The calculated superconducting critical temperature is 
$T_c(\mu_c^*=0)=51$~K for a carrier density of 0.6~electrons/cell. 
In this case, the ratios between the gap and the critical temperature are 
$2 \Delta_0^{\rm FEL}/k_B T_c = 1.76$ and $2 \Delta_0^{\pi}/k_B T_c =3.57$, therefore it appears that the gap on the FEL sheet deviates substantially from the standard BCS behavior. Despite such a deviation our calculated distributions of superconducting gaps are described nicely by BCS curves, as shown by the dotted black lines in Fig.~\ref{fig2}(f).

Interestingly, as we show in Fig.~\ref{fig2}(e) and (f),
if we neglect the anisotropy of the electron-phonon interaction, and solve instead the Eliashberg equations using the isotropic average of the Eliashberg function, then the resulting critical temperatures are severely underestimated, by up to a factor of 2. This finding indicates that the standard approach based on the Allen-Dynes equation~\cite{Allen-Dynes},
employed in all previous studies on graphene, is inadequate for studying superconductivity in this material. This observation is in line with similar analyses performed for superconducting MgB$_2$~\cite{choi2}.

\section{COULOMB EFFECTS}
%\noindent
The results presented so far correspond to an ideal scenario whereby the electron-electron Coulomb
interaction is assumed to have 
no effect on the superconducting pairing. To explore more realistic
situations, we show in Fig.~\ref{fig3} the critical temperatures calculated for several
values of the Coulomb pseudopotential $\mu_c^*$ in the typical range used for carbon 
materials~\cite{si}. As a direct consequence of the weak EPC in graphene,
$T_c$ varies strongly with $\mu_c^*$. The trends shown in Fig.~\ref{fig3} are consistent
with a simple analysis based on the Allen-Dynes equation~\cite{Allen-Dynes} 
at variable $\mu_c^*$ (filled green triangles in Fig.~\ref{fig3}). 
Since typical values for $\mu_c^*$ are in the range 0.10-0.20, we expect 
only $n$-doped graphene to exhibit superconductivity upon doping, with $T_c$ of the order 
of 10~K. 

It would be desirable to perform first principles calculations of the superconducting gap including electron-electron effects as in the density functional theory for superconductors (SCDFT)~\cite{scdft1,scdft2}, however the incorporation of such effects in the ME theory is nontrivial. As a simpler alternative we estimate $\mu_c^*$ using the model screened Coulomb interaction proposed in Ref.~\cite{DasSarma2007}
%Ref.~\onlinecite{DasSarma2007} 
and the double Fermi surface average of Ref.~\cite{Cohen1995}. 
%Ref.~\onlinecite{Cohen1995}. 
Using this approach we estimate $\mu_c^* \simeq 0.16$, 
in good agreement with previous results~\cite{si}. The corresponding critical temperature for $n$-doped graphene is $T_c=13$~K, as indicated by the vertical red line in Fig.~\ref{fig3}(b). 
A detailed discussion of the parameter $\mu_c^*$ in a related system was given in Ref.~\cite{Sanna_CaC6}, where the Coulomb parameter of CaC$_6$ was determined by comparing the superconducting gap obtained from the ME theory and that from the SCDFT. 

From this analysis it is clear that electron-electron interactions are important for
the superconducting pairing in graphene. However, while in the context of the Eliashberg theory such
interactions always tend to weaken the pairing, it is also possible that in proximity
of a VHS novel Coulomb effects may emerge \cite{mcchesney} and co-operate with the electron-phonon
mechanism investigated here. 

\section{CONCLUSIONS}
%\noindent
In conclusion, we report the calculations of the superconducting properties of heavily doped graphene
within the {\it ab initio} anisotropic Migdal-Eliashberg theory. 
Our work highlights the delicate interplay between electron-phonon interactions,
anisotropy, and Coulomb effects in this material, and it demonstrates that simplified approaches
based on isotropic approximations, such as the Allen-Dynes equation, are inadequate in this context.

Our main finding is that, when enough carriers
are injected into graphene so that the Fermi level is aligned with the VHS in the $\pi^*$ manifold,
it should be possible to observe conventional superconductivity
at low temperature, due to the extra pairing associated with the FEL band.
In this case, two distinct superconducting gaps should be clearly observable in tunneling
experiments. 
Given the role of the FEL state in the superconductivity of $n$-doped graphene,
and the well-known sensitivity of the FEL energetics to dielectric screening and quantum
confinement~\cite{margine_nfe,topsakal1,topsakal2}, the superconducting state of graphene is expected 
to be rather delicate, and sensitive to device design and materials preparation. 

We hope that this work will serve as a guideline to experimental research 
in the quest for a superconducting state that remains, to date, elusive.

\section{METHODS}
\label{methodology}
The calculations are performed within the local density approximation (LDA) to density-functional 
theory~\cite{lda1,lda2} and norm-conserving pseudopotentials~\cite{nc1,nc2} using {\tt Quantum-ESPRESSO}~\cite{QE}. 
The valence electronic wavefunctions are expanded in a plane-wave basis set with a kinetic 
energy cutoff of 60~Ry. A graphene layer in isolation is described using a supercell geometry.
The optimized lattice parameter is $a=$2.434~\AA\ and periodic replicas are 10~\AA\ apart.
The electron charge density is computed using a ${\it \Gamma}$-centered Brillouin-zone mesh with 
72$\times$72$\times$1 $\bbk$-points and a Methfessel-Paxton smearing~\cite{mp} of 0.10~eV. 
The dynamical matrices and the linear variation of the self-consistent potential are calculated 
within density-functional perturbation theory~\cite{baroni2001} on the irreducible set of a regular 
12$\times$12$\times$1 $\bbq$-point mesh. The electronic wavefunctions required for the Wannier-Fourier
interpolation within the {\tt EPW} code~\cite{giustino_wannier,EPW} are calculated on a uniform and ${\it \Gamma}$-centered 
$\bbk$-points mesh of size 12$\times$12$\times$1. We considered seven maximally-localized 
Wannier functions~\cite{marzari_rmp,wannier} in order to describe the electronic band structure 
up to 1~eV above the VHS in the conduction band.
Two Wannier functions are $p_z$-like states (one per C atom), three functions are $\sigma$-like 
states localized in the middle of C-C bonds, and two correspond to $s$-like states located directly 
above and below the center of the C$_6$ hexagon. 
We consider two doping scenarios where the Fermi energy matches either one of the
two VHS in the $\pi$ or $\pi^*$ manifolds. This is achieved by using 0.4~electrons per unit cell 
(0.8$\cdot$10$^{15}$~cm$^{-2}$) for $p$-doped graphene, and 0.6~electrons per unit cell
(1.2$\cdot$10$^{15}$~cm$^{-2}$) for $n$-doped graphene.
The isotropic Eliashberg spectral function is defined as:
\begin{equation}
\afo = \frac{1}{\NF N_{\bf k} N_{\bf q}} \sum_{\bbk,\bbkk,\nu} |\gkkl|^2 \deltak \deltakk \delta(\omega-\oql),
\end{equation}
the cumulative EPC is calculated as~\cite{allen_mitrovic}:
\begin{equation}
\lambda(\omega) = 2 \int_{0}^{\omega} d\omega' \alpha^2 F(\omega')/\omega',
\vspace{-0.1cm}
\end{equation}
and the momentum-resolved EPC of each electronic state at the Fermi surface is given 
by~\cite{margine_eliashberg}:
\begin{equation}
\lambda_{\bbk} = \sum_{\bbkk,\nu} \deltakk |\gkkl|^2/\oql.
\vspace{-0.1cm}
\end{equation}
In these expressions $\NF$ represents the density of electronic states per spin at the Fermi level,
$N_{\bf k}$ and $N_{\bf q}$ are the total numbers of ${\bf k}$ and ${\bf q}$ points, 
$\ek$ is the Kohn-Sham eigenvalue referred to the Fermi level, and $\gkkl$ is the screened 
electron-phonon matrix element for the scattering between the electronic states 
$\bbk$ and $\bbkk$ through a phonon with wave vector $\bbq\!=\!\bbkk\!-\bbk$, 
frequency $\oql$ and branch index~$\nu$. Here $\bbk$ and $\bbkk$ indicate both 
the electron wavevector and the band index.
The anisotropic Migdal-Eliashberg equations~\cite{allen_mitrovic, choi1, choi2, margine_eliashberg} 
are given by:
\begin{eqnarray}
\hspace{-0.25cm}
  &&Z(\bbk,i\on) =
   1 + \frac{\pi T}{\NF \on} \sum_{\bbkk n'}
   \frac{ \onp }{ \sqrt{\onp^2+\Delta^2(\bbkk,i\onp)} } \nonumber \\
  &&\qquad\qquad\qquad\times \deltakk \lambda(\bbk,\bbkk,n\!-\!n'), \label{Znorm_surf} \\
 &&Z(\bbk,i\on)\Delta(\bbk,i\on) =
   \frac{\pi T}{\NF} \sum_{\bbkk n'}
   \frac{ \Delta(\bbkk,i\onp) }{ \sqrt{\onp^2+\Delta^2(\bbkk,i\onp)} } \nonumber \\
  &&\qquad\qquad\qquad\times \deltakk \left[ \lambda(\bbk,\bbkk,\!n-\!n')-\mu_c^*\right].\nonumber\\
\label{Delta_surf}
\end{eqnarray}
Here $\on=(2n+1)\pi T$ with $n$ integer are fermion Matsubara frequencies, and $T$ is the absolute
temperature. $Z(\bbk,i\on)$ is the mass renormalization function, $\Delta(\bbk,i\on)$ is the
superconducting gap function, $\lambda(\bbk,\bbkk,\!n-\!n')$ is the momentum- and energy-dependent
EPC, and $\mu_c^*$ is the semiempirical Coulomb parameter. From the superconducting gap function
$\Delta(\bbk,i\on)$ we obtain the gap at real-valued frequencies via Pad\'{e} approximants~\cite{pade1,pade2}.
A detailed discussion of the formalism can be found
%in Ref.~\onlinecite{margine_eliashberg} and Ref.~\onlinecite{suppl}.
in Ref.~\cite{margine_eliashberg}.
For the solution of the Eliashberg equations it is absolutely critical
to use extremely fine Brillouin-zone grids~\cite{margine_eliashberg}. In this study we use
grids containing 400$\times$400$\times$1 $\bbk$-points and 200$\times$200$\times$1 $\bbq$-points, 
respectively. % uniform $\Gamma$-centered grids
The frequency cutoff in the Migdal-Eliashberg equations is set to five times the maximum 
phonon frequency, and the Dirac delta functions 
are smeared using Lorentzian broadenings of 100~meV and 0.5~meV for electrons and phonons, respectively.
All the technical details of the Migdal-Eliashberg calculations
are described extensively in Ref.~\cite{margine_eliashberg}.

F.G. acknowledges support from the European Research Council (EU FP7/ERC grant no. 239578 and
EU FP7/grant no. 604391 Graphene Flagship) and the Leverhulme Trust (Grant RL-2012-001).

\end{document}